# An Attention-Guided Deep Regression Model for Landmark Detection in Cephalograms

Zhusi Zhong[1], Jie Li[1], Zhenxi Zhang[1], Zhicheng Jiao[2] and Xinbo Gao[1(✉)]

[1] School of Electronic Engineering, Xidian University, Xi'an 710071, China
`xbgao@mail.xidian.edu.cn`
[2] Department of Radiology and BRIC, University of North Carolina at Chapel Hill, Chapel Hill, NC 27599, USA

**Abstract.** Cephalometric tracing method is usually used in orthodontic diagnosis and treatment planning. In this paper, we propose a deep learning based framework to automatically detect anatomical landmarks in cephalometric X-ray images. We train the deep encoder-decoder model for landmark detection, which combines global landmark configuration with local high-resolution feature responses. The proposed framework is based on a 2-stage u-net, regressing the multi-channel heatmaps for landmark detection. In this framework, we embed attention mechanism with global stage heatmaps, guiding the local stage inferring, to regress the local heatmap patches in a high resolution. Besides, an Expansive Exploration strategy is applied to improve robustness while inferring, expanding the searching scope without increasing model complexity. We have evaluated the proposed framework in the most widely-used public dataset of landmark detection in cephalometric X-ray images. With less computation and manually tuning, the proposed framework achieves state-of-the-art results.

## 1 Introduction

Cephalometric analysis is a standard tool to quantitatively analyze the human skull and mandible, usually used in maxillofacial surgeries and orthodontic treatments. Cephalometric evaluation is based on some anatomical landmarks on the skull and surrounding soft tissue. Although newer techniques such as cone beam computed tomography (CBCT) begin to apply, due to the high price, the traditional 2D longitudinal section X-ray image of human head is still the most widely used in the cephalometric analysis. No matter which kind of data modality is adopted, the landmarks are still annotated manually, which remains a time-consuming work for an experienced doctor. Moreover, the manual annotation is extremely subjective to observer variability. Because the 2D X-ray images are the projection of the spatial structure which contains anatomical differences across organizations with individual difference, the automatic detection is a challenging problem. Despite the challenges, the identification of the skeletal structure contained in cephalograms is the key to the automatic detection. Therefore, an automatic annotation method would release orthodontists from the time-consuming work and especially avoid the observation errors. Our study concentrates on detecting the widely used 19 landmarks from the 2D radiograph automatically.

**Related work:** More recently, the automatic landmark detection was held as a Grand Challenge at ISBI 2015. The organizers provided the dataset [1] and published the benchmark of the dental radiography analysis algorithms [2]. Ibragimov et al. [3] computerized cephalometry by game-strategy with a shape-based model, Lindner et al. [4] won first place with Random Forest regression-voting method. After that, Lindner et al. [5] expanded their experiments with 4-fold cross-validation on all the data and presented the results with comprehensive experimental analysis.

Deep learning methods have achieved great success in the field of medical image analysis [6-9]. The cascade and hierarchy are the basic idea to improve performance from coarse to fine. Lee et al. [10] applied deep learning method to cephalometric landmark detection for the first time. They trained 38 independent CNN structures to regress the 19 landmarks' $x$- and $y$-coordinate variables separately. As most of the existing landmark detection methods, they need to train a number of models to refine each point on a small scale one by one, which demands massive but inefficient computation.

Different from the traditional coordinate regression methods, deep encoder-decoder methods, such as u-net [11] and fully convolutional networks (FCN) [12], achieve the goal with target transform. In medical landmark detection, by regressing heatmaps for landmarks simultaneously instead of absolute landmark coordinates, Payer et al. [13] transformed the coordinate regression problem to a pixel regression problem and simplified the procedure with multi-layer cascaded deep neural networks. These pixel-to-pixel heatmap regression methods are intrinsically more suitable for landmark detection, they extract the location information from X-ray images, with less divide between data forms than coordinate.

**Contribution:** In this paper, we propose a novel deep learning framework for automatically locating the anatomical landmarks in 2D cephalometric radiographs. The proposed method regress heatmaps of landmarks from coarse to fine in 2 stages, informing global configuration as well as accurately describing local appearance. The Attention-Guide mechanism connects the coarse-to-fine stages, which is similar to [14] but our Attention-Guide mechanism makes effect on several regions simultaneously. The high efficiency of our framework owes to these strategies: (i) our patch-based strategy optimizes the utilization of convolution kernels, to learn the informative feature around landmarks; (ii) the proposed Attention-Guide mechanism acts as an information extractor while inferring and minimizes the proposal region of sliding-window; (iii) with our Expansive Exploration strategy, the framework infers in a large scope, refining local heatmaps without increasing model complexity. The stage-wise training process makes our framework trainable.

## 2 Method

**Overall Framework:** As shown in Fig. 1, the overall framework for landmark detection includes 2 stages, regressing 20-channel heatmaps of landmarks from coarse to fine. The two stages share the same u-net structure (Fig. 1c), but they are assigned with different learning scopes. Stage 1 trains the u-net with the global field, as "global stage", regressing the global heatmaps $H_G$ as landmark configuration. Stage 2 is assigned as "local stage", with patch-based u-net model. Guided by the coarse attention from $H_G$, local stage searches in the proposal regions, regressing the heatmap patches $H_P$ in a high resolution. As shown in Fig. 2, the Expansive Exploration strategy refines each landmark by multiple inference. The predicted coordinates are obtained as the locations of highlights in first 19 channels of heatmaps $H_M$, which is merged from $H_P$.

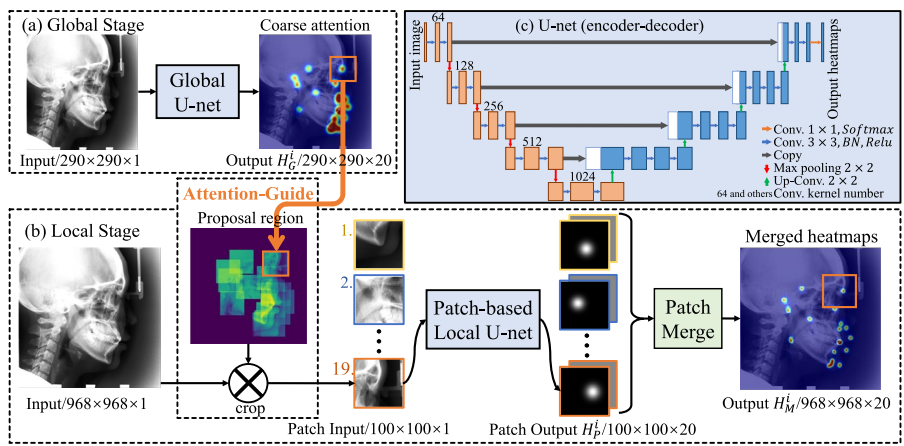

**Fig. 1.** Overall framework of the Attention-Guided deep regression model. (a) Global stage is shown at the top left. (b) Local stage embedded with Attention-Guide is shown in bottom. (c) We illustrate the u-net as encoder-decoder for global u-net and patch-based local u-net.

**Target transform:** Inspired by [13], we convert the coordinate regression to a heatmap regression task. First of all, we represent the abstract coordinates $L$ of 19 landmarks as the 20-channel concrete heatmaps $H$. We model each landmark as a channel heatmap with a 2D Gaussian distribution centered at the landmark. The distribution is normalized to a range of 0 to 1 and the Standard Deviation σ depends the size of distribution. The Correlation Coefficient $\rho$ is 0, to make sure the shape of distributions is circular:

$$G(x,y) = exp\left[-\frac{1}{2\sigma^2}((x-\mu_1)^2 + (y-\mu_2)^2)\right] \quad (1)$$

where $(\mu_1, \mu_2)$ is the center of the distribution. In the circular area of one channel, the pixel values indicate appearing probability of the landmark, so that the distributions can contain the uncertainty which involved in the landmark locations. However, the distributions are much smaller than the outside areas which represent negative class for a channel. We handle this class-imbalance problem similar to [15]. We apply the classification approach to estimate a shared background channel additionally. So that, the 20-channel heatmaps $H$, which represent classes separately, are described as follow:

$$H^i(x,y) = \begin{cases} exp\left[-\frac{1}{2\sigma^2}((x-x_i)^2 + (y-y_i)^2)\right], & i = 1,2,\dots,19 \\ 1 - \sum_{j=1}^{19} H^j(x,y), & i = 20 \end{cases} \quad (2)$$

where heatmap $H^i$ denotes a channel whose distribution is located at the position of landmark $L_i = (x_i, y_i)$, while $i$ is in the range of 1 to 19. And the last channel of heatmaps $H^{20}$ represents the shared background, to ensure the sum of all 20 classes probabilities is 1 for each pixel. The specific variable $\sigma$ is different at stages according to target distribution size. The coordinate regression problem is transformed to a pixel classification task, which achieves goal by regressing the 20-channel heatmaps $H$.

**Global stage and pixel regression:** The global stage takes the entire images as input, and informs the underlying global landmark configuration. We train a modified u-net (Fig. 1c) as the backbone model, followed by a SoftMax activation layer to separate pixel classes probability in channels. Limited by the computational capabilities and the learning ability of the neural network, we have to scale the training data to small size. The output is the 20-channel heatmaps $H_G$, as shown in Fig. 1a (right). The channel-wise highlights indicate the high appearing probability of landmarks. Some areas overlap together on the schematic, it is actually due to compress multi-channel distributions of close landmarks into a plane.

Although, the large size of distributions limits the accuracy of prediction. The convolution kernels cannot distinguish subtle features from low resolution data, and the network cannot regress heatmaps with the small distributions. Besides, the prediction errors increase as sizing back to the original scale. So, we take those highlights on $H_G$ as the coarse attention for local stage, and design a patch-based structure to narrow the learning scope, in order to process data and feature maps in a higher resolution.

**Local stage and Attention-Guided inference:** The local stage with patch-based u-net, guided by the coarse attention, focuses on learning local appearance around landmarks. The patch-based u-net shares the same structure with global stage. But it is trained with the small image patches, which is sampled around ground-truth labels $L_{GT}$ randomly. The local stage learns to regress multi-channel heatmap patches $H_P$ with smaller Gaussian distributions than global stage. So, the local stage u-net informs the high-resolution local features and has better distinguishing ability than the global stage. Our patch-based strategy optimizes the efficiency of local training process, avoiding the negative impact of the areas without landmarks.

The Attention-Guide mechanism is embedded in the local stage inference. We firstly resolve the 19 coarse coordinates which are obtained as the maximum in the first 19

5channels of $H_G$. We set the coarse locations as center of the proposal regions. As shown in Fig. 1b (center), the proposal regions guide the patch selection, by cropping patches in the input image at the corresponding places. Combining with the patch-based strategy, the Attention-Guide acts as an information extractor for local stage, to minimize the proposal region of sliding-window. Local stage takes these image patches as input, regressing 19 heatmap patches $H_P$. Then $H_P$ are normalized and merged to the complete heatmaps $H_M$. As shown in Fig. 1b (right), $H_M$ gather highlights in the small points, those in the overlap regions are refined to smaller and more precise. The 19 predicted coordinates $L_P$ are obtained as the locations of highlights in the first 19 channels of $H_M$. The details are described in the experimental section.

**Expansive Exploration:** The small searching scope of local stage obtains most of landmarks successfully, but the coarse stage dose not guarantee that all landmarks are detected in the proposal regions. To increase the robustness, we propose the Expansive Exploration strategy for the Attention-Guided inference at the local stage, similarly to the overlap-tile strategy in [11]. As shown in Fig. 2, we enlarge the sampling scope and fix the relative position for multiple inference. The expansive proposal regions are the expanded squares centered at coarse locations. The image patches are inputted to local u-net separately, expanding search scope without expansion of the network structure. Overlap margin is controlled by the expand parameter $\varepsilon \in (1,2)$.

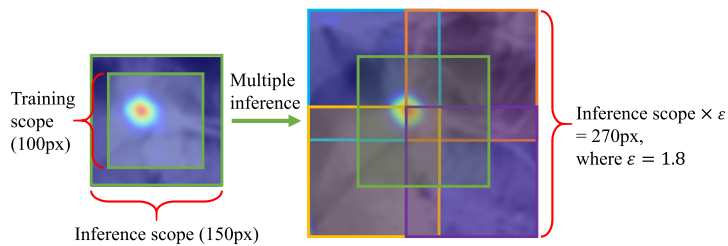

**Fig. 2.** Expansive Exploration strategy for local inferring. We firstly enlarge the single inference scope to 150px (the training scope is 100px), then apply stack searching with fix relative position. 4 regions overlap each other with margin to be a big square, and 1 region places in the center of the expanded square. Here we set the expand parameter $\varepsilon$ to 1.8.

**Heatmap Regression Loss:** Considering the class-imbalance problem, which means the areas as negative class are much larger than those of landmarks, and the small size of distribution target, we add a combination of binary cross-entropy loss (BCE loss) and focal loss [16] as the loss function to balance the cost of background and targets, which is described as follows:

$$L(H, \widehat{H}) = -\frac{1}{N}\sum_{b=1}^{N}\left(\frac{1}{2} \cdot H \cdot \log\widehat{H} + \frac{1}{2} \cdot \alpha_t \cdot (1-H_t)^\gamma \cdot \log H_t\right), \quad (3)$$

$$\text{where } H_t = \begin{cases} \widehat{H} & \text{if } H > 0.01 \\ 1-\widehat{H} & \text{otherwise} \end{cases}$$

where $\hat{H}$ and $H$ denote the predicted heatmaps and the ground-truth heatmaps generated from $L_{GT}$, and $N$ indicates the batch size. The BCE loss plays a major role in evaluating the areas with most background. Then the focal loss tends to mainly finetune the target regions in Gaussian distributions after 60 epochs in our experiments.

## 3      Experiments

This study includes the widely-used public dataset from the Grand Challenge. The dataset contains 400 dental X-ray cephalometric images and the 2 sets of annotations with 19 landmarks from 2 experienced doctors. The data is divided into 3 sets, 150 images for training data, 150 images for Test1 data and 100 images for Test2 data. The resolution of images was 1935×2400 pixels with a pixel spacing of 0.1 mm. We crop the images to squares (1935×1935px) and the annotated y-axis coordinates are subtracted by 465, as shown in Fig. 3c.

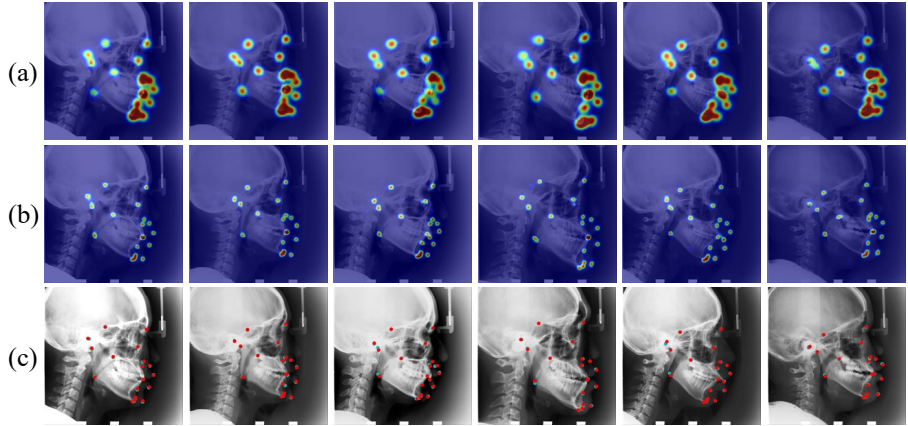

**Fig. 3.** Examples of prediction on testing X-ray images. (a) 1st row shows the channel-compressed $H_G$. (b) 2nd row shows the channel-compressed $H_M$. (c) 3rd row shows the predicted landmarks (red) transformed from $H_M$, and the ground truths (green).

For global stage training, we scale the cropped images and the coordinates by 0.15 times. The global heatmaps $H_G$ are the same size as the scaled images. The Gaussian distributions of $H_G$ are 40-pixel width which are 267-pixel width in the original scale. The global u-net takes the scaled images as input, and it learns to regress the 20-channel coarse heatmaps $H_G$. The channel-compressed results are shown in Fig. 3a.

For local stage training, we down-sample image data by 0.5 times. The original entire heatmaps are as large as the scaled image, with 30-pixel width distribution, whose original width is 60 pixels. The patch-based u-net is trained by randomly sampling an image patch (100×100px) around one landmark a time, to regress the correspond 20-channel heatmap patches $H_P$ cropped from the entire heatmaps. Through numbers of training epochs, the sampler randomly travers all landmarks. The two-stage networks are trained separately with our Heatmap Regression Loss.

While inferring, the patch selection is guided by the coarse attention with our Expansive Exploration strategy. The expansive $H_P$ of each landmark are merged to the $H_M$ by placing at the corresponding location in the expansive proposal regions. In $H_M$, we assume there is no landmark out of patches, so the pixel values of these areas are 0. For overlapping areas, we average the pixel values to raise robustness, receding the artifacts (fake shadow). The first 19 channels of $H_M$ are normalized to a range of 0 to 1 each channel separately, then pass a filter with threshold of 0.5 to reduce the artifacts whose pixel values are less than 0.5. The final 19 coordinates $L_P$ are the mean positions of nonzero pixels each channel separately, they represent the centers of the distributions with high possible of the landmarks.

The mean radial error (MRE, in mm) and the successful detection rate (SDR, in %) are the evaluation indexes of the Grand Challenge. The MRE is defined by $MRE = (\sum_{i=1}^{n} R_i)/n$ where $n$ indicates the number of data and $R$ indicates the Euclidean distance between ground truths $L_{GT}$ and prediction $L_P$. The Std indicates the error's standard deviation in dataset. The SDR shows the percentage of landmarks successfully detected in a range of 2.0 mm, 2.5 mm, 3.0 mm, 4 mm.

We have evaluated the proposed method on 2 experiments. In the challenge, they tested on Test1 data and Test2 data independently, and took the average of two sets of annotations as the ground truth. The comparisons are shown in the first 2 blocks in Table 1. After the challenge, Lindner et al. [5] applied the 4-fold cross-validation experiments on the dataset with all 400 cases, and the ground truths were the annotations from the senior doctor. We follow experiments settings and the results are shown in the third block in Table 1. The Fig. 4 shows the 4-fold cross-validation result of our method.

**Table 1.** Comparison on proposed Deep Regression Model with other approaches

| Test Data | Method | MRE ± Std (mm) | SDR (%) | | | |
|---|---|---|---|---|---|---|
| | | | 2.0 mm | 2.5 mm | 3.0 mm | 4.0 mm |
| Test1 Data | Ibragimov et al. (2015) | 1.84 ± 1.76 | 71.70 | 77.40 | 81.90 | 88.00 |
| | Lindner et al. (2015) | 1.67 ± 1.48 | 74.95 | 80.28 | 84.56 | 89.68 |
| | Ours (stage1) | 1.90 ± 1.17 | 62.41 | 75.63 | 83.82 | 93.72 |
| | Ours (stage2 no Expand) | 1.22 ± 1.42 | 85.38 | 91.19 | 94.21 | 97.27 |
| | **Ours** | **1.12 ± 0.88** | **86.91** | **91.82** | **94.88** | **97.90** |
| Test2 Data | Ibragimov et al. (2015) | | 62.74 | 70.47 | 76.53 | 85.11 |
| | Lindner et al. (2015) | 1.92 ± 1.24 | 66.11 | 72.00 | 77.63 | 87.42 |
| | Ours (stage1) | 2.28 ± 1.72 | 52.53 | 66.00 | 77.58 | 89.53 |
| | Ours (stage2 no Expand) | 1.22 ± 1.42 | 74.42 | 82.42 | 88.11 | 94.63 |
| | **Ours** | **1.42 ± 0.84** | **76.00** | **82.90** | **88.74** | **94.32** |
| 4-fold Cross | Lindner et al. (2016) | 1.20 ± **0.06** | 84.70 | 89.38 | 92.62 | 96.30 |
| | Ours | **1.22** ± 2.45 | **86.06** | **90.84** | **94.04** | **97.28** |

## 4  Conclusion

Our deep learning framework achieves good performance in detecting anatomical landmarks in cephalometric X-ray images. In our framework, the landmark detection task transforms to classification of image pixel. The Attention-Guide and the Expansive Exploitation strategy make sure that the searching scopes is smaller and data resolution is higher with minimum information redundancy. The data augmentation is embedded in the random sampling to avoid overfitting. Our model with higher efficiency but less manual tuning achieves a state-of-the-art result on automatic landmark detection in cephalometric radiograph. Moreover, the encoder-decoder structure which we apply with u-net, is easily transferred to any other model with better performance. And our deep regression model is easily generalized to other landmark detection tasks.

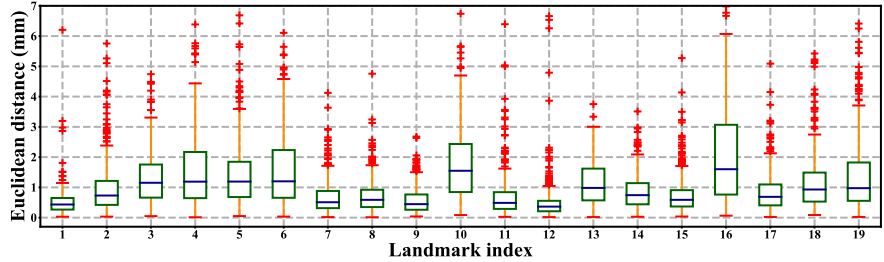

**Fig. 4.** Boxplot of Euclidean distances between predicted landmarks $L_P$ and ground truths $L_{GT}$.

**Acknowledgement.** This work was supported in part by the National Natural Science Foundation of China under Grant 61671339, 61432014 and 61772402, and in part by National High-Level Talents Special Support Program of China under Grant CS31117200001.